# Micro-sized tunable liquid crystal optical filters

CALEB STOLTZFUS,[1] RUSSELL BARBOUR,[2] DAVID ATHERTON,[1] ZEB BARBER[1],*

[1]*Spectrum Lab, Montana State University, Bozeman MT 59717, USA*
[2]*Advanced Microcavity Sensors, Bozeman MT 59715, USA*
*\*Corresponding author: zeb.barber@montana.edu*

**Liquid Crystal Arrayed Microcavities (LCAM) is a new technology for ultra-narrow optical filtering (FWHM ~ 0.1 nm) that uses pico-liter volume Fabry-Perot type optical cavities filled with liquid crystal for tuning. LCAMs are sub-nm spectral resolution filters, which utilize well-established laser writing, thin film deposition, and wafer manufacturing techniques. These filters are compact, robust, and inexpensive. Compact, high-resolution optical filters have applications including biomedical imaging, chemical detection, and environmental monitoring. Here we describe the LCAM design and initial performance metrics.**

Tunable optical filters come in many shapes, sizes, and complexities – each addressing the unique requirements of disparate applications that need various bandwidths, tuning speeds, and spectral tuning ranges. Piezoelectrically driven, air-gap Fabry-Perot filters [1–6], microcavity filters [7–11], birefringent filters [12–14], and liquid crystal based devices [15–19] have been used to fill the various requirements of different applications.

Liquid crystal tunable filters (LCTFs) have been used for Raman imaging, chemical detection, hyperspectral imaging, and many other applications [5,17–21]. LCTFs are typically either Fabry-Perot or Lyot-type filters. The use of liquid crystal to tune the peak transmission wavelength allows for robust, non-mechanical, low voltage controlled tunable filtering of input light. Commercially available LCTFs cover a broad range of wavelengths. However, they typically have limited spectral resolution (limited by the achievable parallelism of the mirrors and no lateral confinement) and can be relatively large in size. These filters have a FWHM > 5 nm tunable bandpass and transmission window and exhibit maximum transmission, T = 5% - 60%, for polarized light.

Lyot-type filters use multiple birefringent plates sandwiched between parallel polarizers to achieve spectral selection [13,17]. Using liquid crystal in the birefringent plates allows for electrical tuning of the transmission wavelength. Lyot-type filters typically have high attenuation due to the use of stacked polarizers [13]. These filters have a broad acceptance angle, requiring no mode coupling, which makes them ideally suited for many applications.

Fabry-Perot type microcavity filters constructed from two closely spaced mirrors (spaced with an air-gap < 2μm) are also used to filter broadband light. Recent advances in microcavity technology have been primarily driven by the need for high finesse and small mode volume cavities in the field of atomic physics and solid-state cavity QED [6]. Several groups have constructed open-access optical microcavities formed by a microscopic curved mirror and a planar mirror combination that demonstrates a cavity finesse that is limited only by the quality of the mirror coatings [1,3–5]. A cavity finesse of more than 100,000 has been demonstrated [22]. The distance between the mirrors determines the transmission wavelength, which is tuned using piezoelectric actuators. These filters provide very narrow transmission bandwidths, Δλ << 1 nm. The use of piezos to tune the air gap length introduces challenges for a useful commercial device such as drift, hysteresis, and non-linearity. When used in a mechanically isolated and thermally controlled cryogenic environments, the majority of piezo issues are resolved and these filters have demonstrated excellent performance [4].

Many applications require a robust, room temperature, sub-nm resolution, and continuously tunable filter. By combining liquid crystal tuning with Fabry-Perot type microcavities, the filters described here achieve all of these specifications [23]. Our current design, using mirrors with a peak reflectivity of 99.4%, provides continuously tunable transmission that is less than 0.3 nm FWHM over a 200 nm (550 nm-750 nm) bandwidth and with a free-spectral range of 20 nm – 70 nm. With higher reflectivity mirrors, transmission widths of 10s of picometers can be achieved. A 10 to 100 μm diameter curved coupling mirror on one of the Fabry-Perot mirror surfaces forms a stable pico-liter sized optical cavity. A voltage across the liquid crystal between the two mirrors tunes the index of refraction of the liquid crystal, which tunes the peak transmission wavelength of the filter for linearly polarized light, whose polarization is aligned with the liquid crystal molecules, while the peak transmission for orthogonally polarized light remains fixed. This means the tunable transmission is only effective for one polarization and external polarization optics are required to filter out or address the orthogonal polarization. The use of a monolithic, fixed-gap design reduces the susceptibility of the filter to noise from vibrations and temperature changes.

In this paper we discuss the development and characterization of our liquid crystal arrayed microcavities (LCAM). An LCAM consists of a 2-D array of many narrowband tunable microcavity filters. Each cavity can be used as a single element or combined into an interconnected spatial array. These filters enable highly miniaturized spatial / spectrally resolving devices for material characterization, biomedical imaging, hyperspectral microscopy, and other applications.

**Fig. 1**(A) shows the basic schematic of a single microcavity channel of an LCAM. A miniature (10 – 100 μm) diameter depression is laser written into a fused silica substrate. A conductive Indium Tin Oxide (ITO) and a high reflection (>99%) multi-layer dielectric mirror coating at the desired wavelength is deposited on top of this substrate. A stable Fabry-Perot cavity is formed with a second flat mirror with the same ITO and mirror coating. The mirrors are spaced by a thin (L > 1 μm) gap, which is filled with a nematic liquid crystal oriented with a very thin (< 100 nm) alignment layer. The ITO layers allow low voltage control of the LC tuning layer. **Fig. 1**(B) shows a completed LCAM filter coupled to a fiber with bulk coupling optics. A large array of these channels could be integrated with CMOS or CCD detectors to rapidly collect both spatial and spectral information from an illuminated object or biological sample.

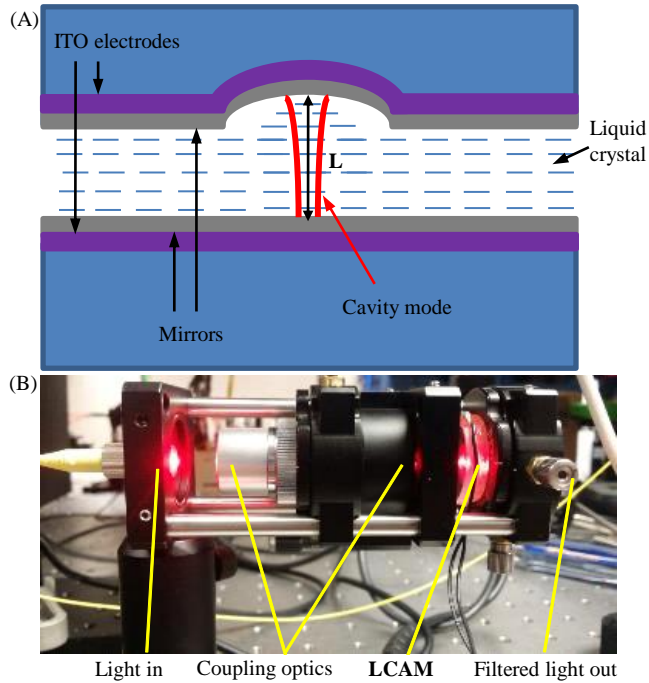

**Fig. 1. (A)** LCAM schematic showing a conceptual design of a single optical microcavity (not to scale). The cavity medium contains a homogeneous planar aligned nematic liquid crystal layer (blue dashes) allowing the effective cavity length to be tuned. **(B)** Current device undergoing benchtop testing. The monolithic compact package is low cost, highly ruggedized, and operates well at room temperatures.

A flow chart of the fabrication process is shown in **Fig. 2**. We start by ablating shallow arrays of craters onto 4" fused silica wafers (University Wafer Part #: U01-131126-11 JGS2, Ra < 1 nm). The ablation process melts the fused silica and surface tension makes the surface of curved mirror atomically smooth. The depth of the craters was varied from 0.2 μm up to 6 μm, with a corresponding radius of curvature of 1300 μm down to 50 μm. The craters are ablated using the 10.6 μm line of a tightly focused $CO_2$ laser (Synrad, FSV30SFG). The $CO_2$ laser is focused onto the wafer using a ZnSe lens (f = 15mm). The intensity of the white light emitted by the plasma generated in the ablation process is used in a feedback circuit to control the laser energy absorbed by the substrate, which is proportional to the depth of the crater. The laser ablation for a single crater occurs in less than 50 μs. Arrays of craters are ablated using an automated LabVIEW program, where computer controlled stages (Zaber, T-LS28E) move the wafer to the position of each crater. Ablating a 10 by 10 array of 220 μm spaced craters takes only a couple minutes – limited mostly by the motion of the stages – which could be greatly improved. Many such arrays can be written on a single wafer. Half of the wafer area, or separate wafers, are reserved (no ablated craters) to form the flat portion of the microcavities.

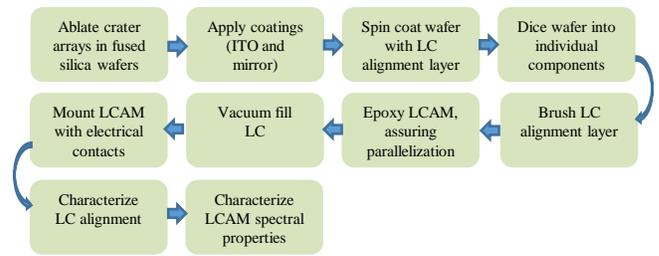

**Fig. 2.** Flowchart of the LCAM manufacturing steps.

Once multiple arrays of craters have been ablated, all the wafers were coated by Optical Filters Source, LLC in Pflugerville, Texas. First a 180 nm thick layer of indium tin oxide (ITO) is deposited. The conductive ITO layer allows for voltage control of the LC. In some cases, the ITO layer is patterned to allow simultaneously addressing craters from the same array with different voltages. Placing the optically lossy ITO layer underneath the high reflectivity mirror eliminates this as a source of optical loss in the Fabry-Perot cavity and only slightly increases the required tuning voltage. After the ITO has been deposited, seven pairs of $TiO_2$/$SiO_2$ mirror coatings are deposited yielding a peak reflectivity of 99.4% centered at 615 nm. The high reflectivity mirror coating determines many of the properties of the final LCAM device including wavelength coverage and resolution of the Fabry-Perot resonances. Broadband dielectric mirror designs can be used to increase the wavelength coverage, but require more layers, which can increase the losses due to the mirrors and effect the filter tuning due to the dispersion characteristics of the coating.

Once the ITO and mirror coatings have been deposited, the wafers without craters are spin coated with a ~100 nm thick layer of 2% polyvinyl alcohol (PVA) dissolved in DI water, which forms the LC alignment layer. After this the wafers with craters are diced into 9.5 mm by 9.5 mm square pieces, and the wafers without craters are diced into 12.7 mm by 12.7 mm square pieces. The difference in size between the two mirror pieces allows for easy access to the conductive ITO layer after LCAM assembly.

To align the liquid crystal, we gently drag lens tissue across the PVA layer 100s of times. Once the alignment layer has been brushed we fix the two mirror surfaces together using UV cure epoxy (Norland UV sealant 91 epoxy) mixed with ~1 μm diameter silica spacer beads (Cospheric, SiO2MS-1.8 0.961um). Before the epoxy is cured a custom jig is used to assure that the mirror pieces are parallel by centering the Newton rings. Typically, this method yields a sufficiently parallel gap over the small area of an individual optical cavity. To fill the gap between the mirrors with LC the LCAM unit is placed in a vacuum chamber and a drop of LC is placed on the edge of the LCAM, where it can drain into the gap. After the LC is placed on the LCAM, the vacuum pump is turned on as soon as possible. The vacuum chamber pumps down to a modest vacuum much faster than the LC fills the cavity, which ensures that air bubbles are minimized in the final LC fill. For the LCAMs described here either 5CB or E7 liquid crystal were used. The LC alignment and LC filling is checked with a polarization microscope. Microscope images of the craters and liquid crystal layer are shown in **Fig. 3**. For some LCAMS the unit is heated to beyond the clearing temperature of the LC for a few minutes to achieve good LC alignment.

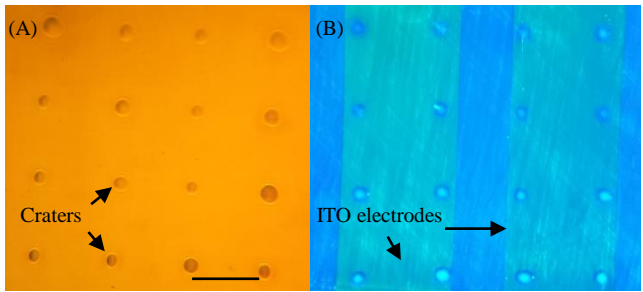

**Fig. 3. (A)** Reflected light microscope image of the craters in an LCAM. **(B)** Polarization microscope image showing excellent LC alignment over an array of LCAM channels. The PVA alignment layer was fabricated as described in the text. The greenish strips are ITO electrodes that enable independent tuning of different LCAM channels, and the circles are the craters in the mirror surface. The scale bar is 200 μm.

**Fig. 4** (A) and (B) show the characteristic transmission of a single LCAM optical cavity with broad white LED light illumination. **Fig. 4** (C) shows the characteristic transmission of the same optical cavity with 632 nm HeNe laser illumination. In both cases, the illumination was coupled to a single optical cavity using a 40x, 0.45 NA microscope objective. A 60 kHz square wave with a varying peak-to-peak voltage ($V_{pp}$) is applied to the ITO layers to control the orientation of the LC molecules. As the voltage across the LC is varied from 0 to 10 volts the peak transmission tunes to shorter wavelengths. Each horizontal row in **Fig. 4** (A) represents a full spectrum of the LCAM transmission corresponding to each LC voltage. The transmission spectrum when $V_{pp}$ = 2.54 V is shown in **Fig. 4** (B). The reflectivity of the mirror coatings, although not perfectly spectrally flat, is at least greater than >90% within the wavelength region between 550 nm and 750 nm, allowing the transmission to be tuned anywhere within this 200 nm wavelength range. This range can be set by changing the mirror coatings. The free spectral range (FSR) and the FWHM of the transmission peaks can be adjusted by changing the spacing between the mirrors. As can be seen in the bottom portion of **Fig. 4** (A), below ~1 $V_{pp}$, the liquid crystal does not respond and no tuning occurs. Above the 1 $V_{pp}$ threshold the transmission of each peak can be tuned across 67 nm.

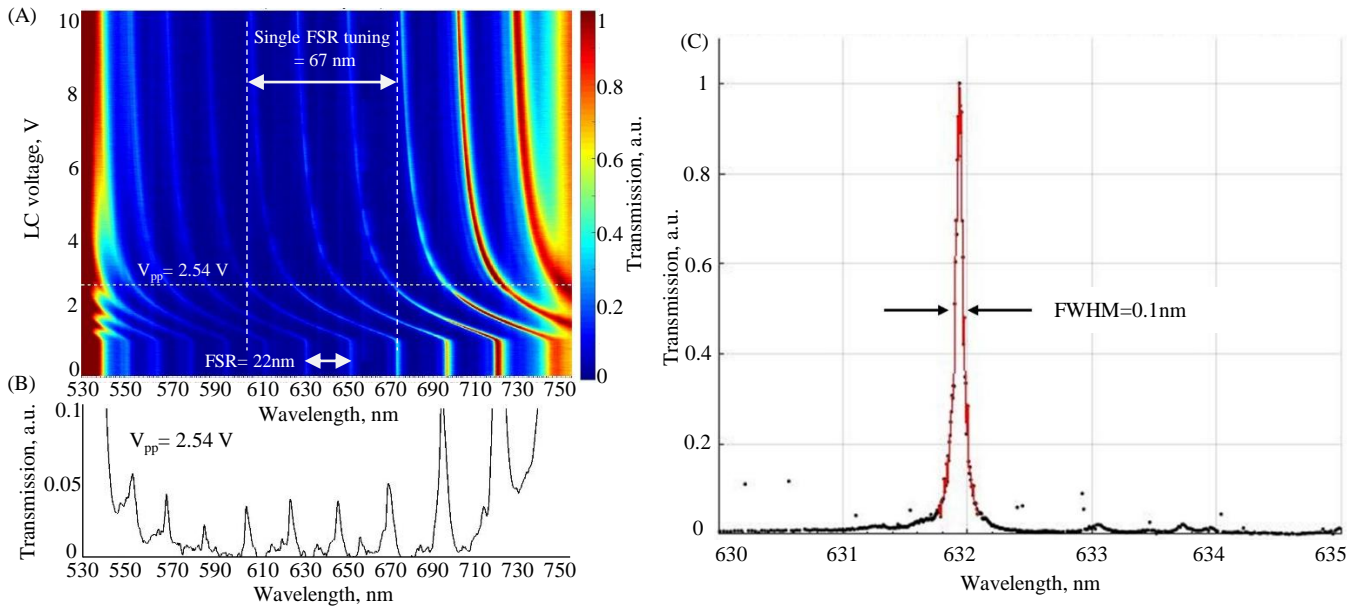

**Fig. 4. (A)** Transmission of a single LCAM cavity, from LCAM 3, with spectrally broad illumination. Cavities are tunable over a wide wavelength range. **(B)** Spectrum of the LCAM transmission corresponding to Vpp = 2.54 V. The limited resolution of the spectrometer (2 nm resolution, Stellarnet, GREEN-Wave) makes the transmission peaks appear broader and dimmer than they should. **(C)** Transmission of the same LCAM channel with 632 nm HeNe illumination, recorded with high spectral resolution.

**Fig. 5** shows the performance of multiple optical cavities from different LCAM units. The finesse refers to the ratio of the FSR divided by the bandwidth (FWHM) of a single transmission peak. Finesse is independent of the FSR of the cavity and is a measure of the optical scattering losses in the cavity. The 632 nm line of a HeNe laser was used to measure the FWHM. The laser was coupled to a microcavity and the transmission was measured as a function of $V_{pp}$. Wavelength versus $V_{pp}$ data, as is shown for a single cavity in **Fig. 4** (A), was used to calibrate the conversion from $V_{pp}$ to wavelength. In **Fig. 4** small transverse mode transmission peaks can be observed. These are suppressed with good alignment to the TEM00 mode of the optical cavity. In applications where good alignment cannot be achieved, deconvolution of the characterized transmitted spectrum may be required.

Each LCAM unit has 100 optical cavities in a 10 × 10 array. Multiple units were made with either E7, 5CB, or no liquid crystal. The uniformity of the liquid crystal alignment layer varied when different brushing techniques were used. The most successful technique involved lightly brushing the PVA layer with lens cleaning tissue 100s of times, as is described previously. This technique was used for all LCAMs described here. The two LCAM units with no LC used piezoelectric mounts to scan the physical distance between the mirrors.

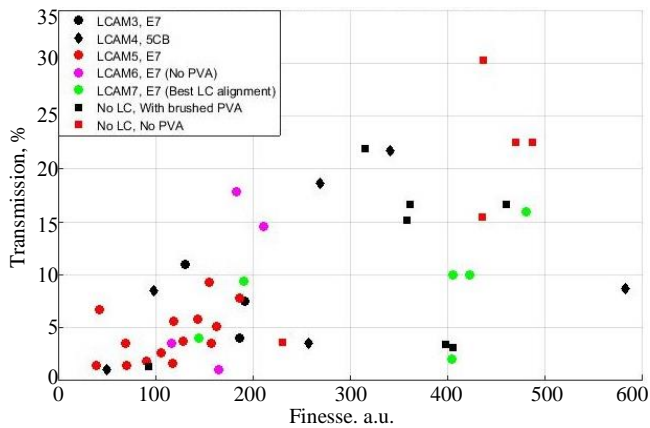

**Fig. 5.** Transmission versus finesse plot for a selection of optical cavities from different LCAM units.

The data shown in **Fig. 5** suggest that the LC could somewhat reduce the quality of the resonator. However, it appears that as more LCAMs were assembled their properties improved, with some achieving near the theoretical finesse of 522 limited by the 99.4% reflectivity of the cavity mirrors. This is likely due to the fact that the final steps of assembling the LCAM units is done by hand and takes some skill, which improved with practice. There is significant variation of both the finesse and the transmission from cavity to cavity. The most likely cause of this is particulate matter left in the craters from the ablation process, handling in non-cleanroom environments, and local defects in the dielectric mirror and PVA coatings. Some craters appeared to have very little contamination and defects and some had lots. Improvements to the ablation and manufacturing processes should yield more uniform cavity performance.

In summary, the development and characterization of initial LCAMs yielded sub-nm resolution, room temperature, tunable filters. Future improvements to the LCAM design will increase the operation range for a wider accessible wavelength region and increased uniformity of individual cavities across arrays. We aim to expand LCAM spectral coverage to the near infra-red (NIR) during the next phase of development. The long-term vision for the LCAM technology is a low-cost, integrated, and versatile hyperspectral / ultra-spectral imager, which will have high spectral resolution, a low cost-to-performance ratio, and be ultra-compact. These miniaturized filters enable devices for material characterization, biomedical imaging, hyperspectral microscopy, and other applications.

**Funding.** National Science Foundation (NSF) (1548568); Montana Board of Research and Commercialization Technology (MBRCT) (#16-50-001); Air Force STTR FA8650-15-M6656 PHASE I and II.

**Acknowledgment.** The authors would like to thank Dr. Daniel Farkas and Dr. Nicholas Booth of Spectral Molecular Imaging, Inc., and Dr. Krishna Rupavatharam for many useful conversations. We would also like to thank Frank Calcagni from Optical Filter Source Inc., Tom Baur from Meadowlark Optics, and Dr. Hiroshi Yokoyama from Kent State University for useful advice on fabrication and liquid crystal alignment.

## References

1. R. J. Barbour, P. A. Dalgarno, A. Curran, K. M. Nowak, H. J. Baker, D. R. Hall, N. G. Stoltz, P. M. Petroff, and R. J. Warburton, "A tunable microcavity," J. Appl. Phys. **110**, 053107 (2011).
2. D. Hunger, C. Deutsch, R. J. Barbour, R. J. Warburton, and J. Reichel, "Laser micro-fabrication of concave, low-roughness features in silica," AIP Adv. **2**, 012119 (2012).
3. D. Hunger, T. Steinmetz, Y. Colombe, C. Deutsch, T. W. Hänsch, and J. Reichel, "A fiber Fabry–Perot cavity with high finesse," New J. Phys. **12**, 065038 (2010).
4. L. Greuter, S. Starosielec, D. Najer, A. Ludwig, L. Duempelmann, D. Rohner, and R. J. Warburton, "A small mode volume tunable microcavity: Development and characterization," Appl. Phys. Lett. **105**, 121105 (2014).
5. M. Kuznetsov, M. Stern, and J. Coppeta, "Single transverse mode optical resonators," Opt. Express **13**, 171–181 (2005).
6. K. J. Vahala, "Optical microcavities," Nature **424**, 839–846 (2003).
7. A. Poon, F. Courvoisier, and R. Chang, "Multimode resonances in square-shaped optical microcavities," Opt. Lett. **26**, 632–634 (2001).
8. Z. Di, H. V. Jones, P. R. Dolan, S. M. Fairclough, M. B. Wincott, J. Fill, G. M. Hughes, and J. M. Smith, "Controlling the emission from semiconductor quantum dots using ultra-small tunable optical microcavities," New J. Phys. **14**, 103048 (2012).
9. G. Ctistis, A. Hartsuiker, E. van der Pol, J. Claudon, W. L. Vos, and J.-M. Gérard, "Optical characterization and selective addressing of the resonant modes of a micropillar cavity with a white light beam," Phys. Rev. B **82**, (2010).
10. C. Reese, C. Becher, A. Imamoğlu, E. Hu, B. D. Gerardot, and P. M. Petroff, "Photonic crystal microcavities with self-assembled InAs quantum dots as active emitters," Appl. Phys. Lett. **78**, 2279–2281 (2001).
11. P. R. Dolan, G. M. Hughes, F. Grazioso, B. R. Patton, and J. M. Smith, "Femtoliter tunable optical cavity arrays," Opt. Lett. **35**, 3556–3558 (2010).
12. J. Evans, "Solc birefringent filter," J Opt Soc Am **48**, 142–145 (1958).
13. O. Aharon and I. Abdulhalim, "Liquid crystal Lyot tunable filter with extended free spectral range," Opt. Express **17**, 11426–11433 (2009).
14. G. Shabtay, E. Eidinger, Z. Zalevsky, D. Mendlovic, and E. Marom, "Tunable birefringent filters—optimal iterative design," Opt. Express **10**, 1534–1541 (2002).
15. S. Isaacs, F. Placido, and I. Abdulhalim, "Investigation of liquid crystal Fabry–Perot tunable filters: design, fabrication, and polarization independence," Appl. Opt. **53**, H91 (2014).
16. J. S. Patel, M. A. Saifi, D. W. Berreman, C. Lin, N. Andreadakis, and S. D. Lee, "Electrically tunable optical filter for infrared wavelength using liquid crystals in a Fabry–Perot étalon," Appl. Phys. Lett. **57**, 1718–1720 (1990).
17. H. R. Morris, C. C. Hoyt, P. Miller, and P. J. Treado, "Liquid crystal tunable filter Raman chemical imaging," Appl. Spectrosc. **50**, 805–811 (1996).
18. M. Abuleil and I. Abdulhalim, "Narrowband multispectral liquid crystal tunable filter," Opt. Lett. **41**, 1957 (2016).
19. I. August, Y. Oiknine, M. AbuLeil, I. Abdulhalim, and A. Stern, "Miniature Compressive Ultra-spectral Imaging System Utilizing a Single Liquid Crystal Phase Retarder," Sci. Rep. **6**, 23524 (2016).
20. D. L. Farkas and D. Becker, "Applications of spectral imaging: detection and analysis of human melanoma and its precursors," Pigment Cell Res. Spons. Eur. Soc. Pigment Cell Res. Int. Pigment Cell Soc. **14**, 2–8 (2001).

21. A. O. H. Gerstner, W. Laffers, F. Bootz, D. L. Farkas, R. Martin, J. Bendix, and B. Thies, "Hyperspectral imaging of mucosal surfaces in patients," J. Biophotonics **5**, 255–262 (2012).
22. C. J. Hood, H. J. Kimble, and J. Ye, "Characterization of high-finesse mirrors: Loss, phase shifts, and mode structure in an optical cavity," Phys. Rev. A **64**, (2001).
23. R. Barbour and Z. Barber, "Microcavity array for spectral imaging," U.S. patent WO2016057125A1 (April 2016).